\documentclass[%
 aip,
 amsmath,amssymb,
 reprint,%
]{revtex4-1}
\usepackage[english]{babel}

\usepackage{graphicx}% Include figure files
\usepackage{dcolumn}% Align table columns on decimal point
\usepackage{bm}% bold math
\usepackage[utf8]{inputenc}
\usepackage[T1]{fontenc}
\usepackage{mathptmx}
\usepackage{etoolbox}

\makeatletter
\def\@email#1#2{%
 \endgroup
 \patchcmd{\titleblock@produce}
  {\frontmatter@RRAPformat}
  {\frontmatter@RRAPformat{\produce@RRAP{*#1\href{mailto:#2}{#2}}}\frontmatter@RRAPformat}
  {}{}
}%
\makeatother
\begin{document}

\preprint{AIP/123-QED}

\title[Analysis of Performance Limits in Current-Matched Tandem Solar Cells]{Analysis of Performance Limits \\ in Current-Matched Tandem Solar Cells}% Force line breaks with \\% Force line breaks with \\
\author{Anupam Yedida}
% \author{B. Author}%
%  \email{Second.Author@institution.edu.}
\affiliation{ 
  Centre for Nanoscience and Engineering, Indian Institute of Science, Bengaluru 560012, Karnataka, India%\\This line break forced with \textbackslash\textbackslash
}%

\affiliation{%
  Department of Electrical Engineering, 
  Indian Institute of Technology Palakkad, 
  Kanjikkode, Palakkad 678623, Kerala, India.%\\This line break forced% with \\
}%
\author{Revathy Padmanabhan}
 % \homepage{http://www.Second.institution.edu/~Charlie.Author.}
\affiliation{%
  Department of Electrical Engineering, 
  Indian Institute of Technology Palakkad, 
  Kanjikkode, Palakkad 678623, Kerala, India.%\\This line break forced% with \\
}%

\email{anupamyedida@iisc.ac.in; revathyp@iitpkd.ac.in}

\date{\today}% It is always \today, today,

\begin{abstract}
Tandem solar cells are at the forefront of extending the efficiency limits of solar cell technology. Among two-terminal tandems, current-matched (CM) tandem solar cells are of particular interest, owing to their relative ease of fabrication. Though CM tandems have been extensively studied, an analytical framework based on the detailed balance for \(N\)-layer CM tandems with area de-coupled subcells has not been addressed. Current matching constraints can be alleviated with the appropriate addition of area de-coupled subcells/modules at each layer. In this work, we propose an analytical framework for modeling the performance limits of an \(N\)-layer CM tandem solar cell with subcells across layers. Radiative coupling among cells is taken into account. Analytical expressions for the optimal number of subcells across layers are presented. Additionally, we investigate the impact of bandgap mismatches on efficiency, a critical factor in real-world fabrication due to material and processing variations, demonstrating that subcells enhance robustness against such imperfections. Our work provides useful design guidelines for designing and estimating the performance limits of advanced solar cell architectures and can be extended to voltage-matched and bifacial devices as well.
\end{abstract}

\maketitle
\section{Introduction}\label{Section: Introduction}
The transition to renewable energy sources is of paramount importance in the global strategy to combat climate change and achieve sustainability. At the forefront of this transition is photovoltaic (PV) technology, which has seen rapid growth in both research and deployment. They are pivotal in reducing the carbon footprint and curbing our dependence on fossil fuels \cite{globalProgress23,sharma2015solar,ajayan2020review}. The Paris Agreement and subsequent policy initiatives like the European Green Deal have set ambitious targets to reduce carbon emissions \cite{breyer2019solar, kougias2021role}. A recent study by Ernst \& Young claims that the Levelized Cost of Electricity (LCOE) for PV is now \(29\%\) lower than the cheapest fossil fuel alternative \cite{ey2023energy}. In this context, advancing the efficiency and effectiveness of solar technologies is not just a scientific challenge but a necessity for meeting these global environmental goals as well. Emergent technologies such as the field of agrophotovoltaics are a promising avenue of research as incorporating solar energy harvesting and agricultural production within the same area optimizes land usage \cite{pr11030948}. 

The technological evolution of PV cells from traditional single-junction (SJ)  to advanced multi-junction (MJ) configurations, also referred to as tandems, is reflective of the industry's response to the dual challenges of efficiency and economic viability. In PV technology, individual SJ cells, owing to their limited size, each produce a relatively small amount of power. To address the need for larger energy outputs and practical utility, numerous PV cells are interconnected, forming expansive installations known as solar farms \cite{Sindhu_Nehra_Luthra_2017, Khan_Patel_Asadpour_Imran_Butt_Alam_2021, Jahangir_Patel_Asadpour_Khan_Alam_2024}. These installations cover extensive areas and are composed of arrays of PV modules, with each module typically containing \(60\) cells connected in series to generate voltages suitable for practical use \cite{yousuf2022cell, bergveld2012module, bowersox2018design}. 
%This arrangement enables processes such as radiative coupling, which notably augments the total power output \cite{strandberg2015detailed}.

As SJ solar cells have been approaching the Shockley-Queisser (SQ) limit \cite{Shockley_Queisser_1961}, it has prompted the exploration of different designs, such as different tandem configurations, which have the potential to capture a broader spectrum of solar radiation \cite{Brown_Green_2002,Bremner_Levy_Honsberg_2008,lal2017perovskite,geisz2020six,xu2020triple,france2015design,guter2009current}. Therefore, improving the performance of tandem solar cells is of particular interest as they have the potential for higher efficiencies and power outputs, due to their increased utilization of the solar spectrum \cite{meillaud2006efficiency}. The modeling of MJ tandems has been extensively studied \cite{geisz2015generalized,benaicha2020simulation,restat2024optoelectrical,shrivastav2024design,walker2016nonradiative,walker2015impact}. Current-matched (CM) tandem (two-terminal tandem) cells, in particular, are popular owing to their relative ease of fabrication in contrast to unconstrained and voltage-matched (VM) configurations. However, their performance is constrained by requiring the current densities to be the same across the layers, limiting the current density to the cell with the lowest current density. 
%Incorporating subcells can equalize current densities across the layers, presenting two primary benefits: the easing of electrical constraints (achieving current matching) and increased versatility in the selection of bandgap combinations to achieve higher efficiencies.

\begin{figure*}[t]
    \centering
    \includegraphics[scale=0.9]{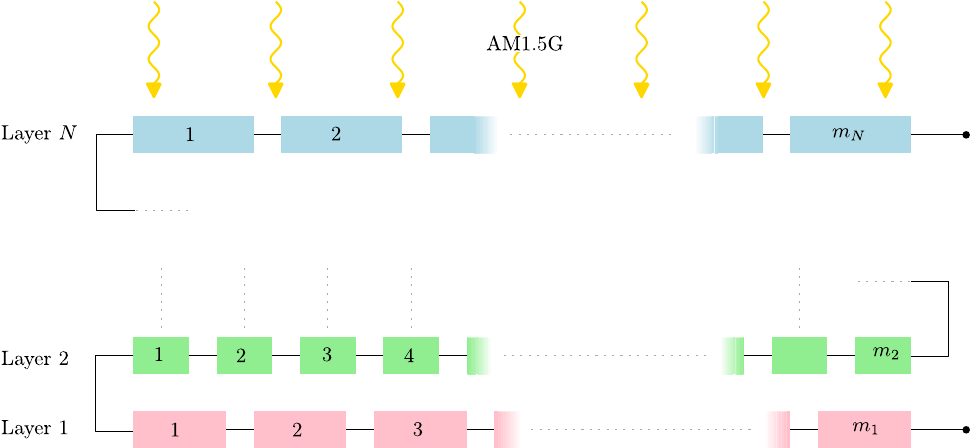}
    \caption{Schematic of a typical device configuration of an $N$-layer CM tandem cell with subcells.}
    \label{fig:device_structure}
\end{figure*}

Incorporating area de-coupled subcells within each layer \cite{strandberg2015detailed,schulte2018string, wagner2024bandgap} helps in alleviating this constraint. These subcells are connected in series across each layer. For CM tandems, these layers are connected in series, and for VM tandems, they are connected in parallel. Incorporating subcells alleviates the current-matching restriction by ensuring that the currents in all the layers are equal. These area de-coupled subcells have been experimentally fabricated in various studies \cite{geerligs2021experimental,alberi2018experimental,mantilla2017monolithic,bobela2017demonstration}. Wagner et al. have used detailed balance to model the efficiency limits for a two-layer VM tandem solar cell \cite{wagner2024bandgap}. Incorporating subcells broadens the spectrum of bandgaps that can achieve high efficiencies, providing greater flexibility in bandgap choices for the layers, although it does not alter the maximum efficiency achievable.
% However, in reality, achieving the exact bandgaps required for optimal current matching is challenging due to material imperfections and process variability. Our study analyzes the impact of these bandgap mismatches on device performance, and we show that tandems with subcells perform better than without subcells when subject to mismatches. 

Achieving the desired bandgap in solar cells can be challenging due to material and processing variations \cite{miah2024band,Brennan_Draguta_Kamat_Kuno_2018,Tang_VanDenBerg_Gu_Horneber_Matt_Osvet_Meixner_Zhang_Brabec_2018, Yu_Sun_Zhang_Zhang_Shen_Wang_2021}. A mismatch in the bandgap can lead to a significant loss in efficiency. This issue is exacerbated in CM tandems, as a bandgap mismatch in both top and bottom layers can lead to significant disruption to the current-matching requirement, thus leading to reduction in overall efficiency of the tandem. 

While analytical models based on detailed balance describing the performance limits for VM tandems have been studied \cite{strandberg2020analytic}, similar analytical frameworks for CM devices with subcells appear to be lacking. 
In this study, we present an analytical framework that assesses the performance limits of \(N\)-layer CM tandems incorporating area de-coupled subcells based on the thermodynamic limit. We make use of the detailed balance principle \cite{klein1955principle} to arrive at the thermodynamic limit. Radiative/Luminescent coupling between layers is accounted for as well. Additionally, we study the impacts of bandgap mismatch on the CM tandems without and with subcells.

This paper is organized as follows: after describing the device configuration in the next section, we expand on our modeling approach and present our calculations and analysis in the subsequent sections. Finally, we discuss the scope and limitations of this approach and furnish our concluding remarks.

\section{Device Model}\label{section: Device_model}

\subsection{Device Structure}\label{sub_section: Device_strucure}
The schematic of an $N$-layer CM tandem cell is shown in Fig. \ref{fig:device_structure}; $N$ is the number of layers, and $m_i$ is the number of subcells/modules in layer $i$. We have assumed the presence of a back reflector at the bottom of layer $1$. Layers are connected in series with layer \(i\) consisting of \(m_i\) subcells. The area between successive cells in a given layer is considered to be negligible, as we are interested in ascertaining the maximum possible performance limit.

\section{Modeling Approach and Assumptions}\label{section: Modeling Framework and Assumptions}

\subsection{Model Assumptions}\label{sub_section: Model_assumptions}
As discussed in Section \ref{Section: Introduction}, our model investigates the performance limits of CM tandem solar cells by ascertaining the thermodynamic limits using the detailed balance approach.  

\begin{figure}[h]
    \centering
\includegraphics{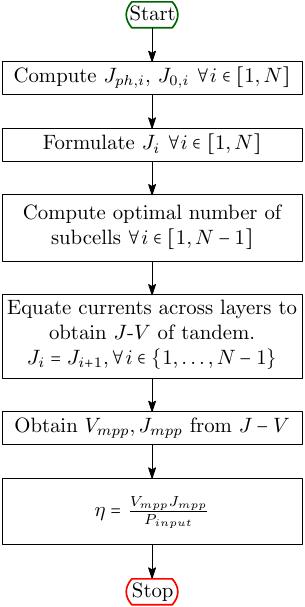}
    \caption{Flowchart summarizing our framework. MPP refers to the maximum power point. The current density - voltage characteristics is denoted by \(J\)-\(V\)  and the voltage and current density corresponding to the MPP are given by \(V_{mpp}\) and \(J_{mpp}\).}
    \label{fig: flowchart}
\end{figure}

\textit{Detailed balance framework:} (a) Full photon absorption for photons with energy greater than or equal to the bandgap energy (\(h\nu \ge E_g\)). (b) Both internal and external quantum efficiencies (IQE and EQE) are assumed to be 100\%. (c) The model incorporates radiative/luminescent coupling between adjacent layers (specifically, between layers \(i\) and \(i+1\)), focusing on the coupling from higher bandgap to lower bandgap layers. (d) The impacts of series and shunt resistances along with any parasitic losses are ignored, as the detailed balance framework addresses the maximum theoretical efficiency.

\subsection{Modeling Framework}\label{sub_section: Modeling_framework}
Our work assumes radiative recombination as the exclusive loss mechanism, and that the solar cell acts as a black body that inherently emits radiation. The current density lost due to this process in a SJ cell of band gap $E_g$ emitted towards both hemispheres is given by \(J_{dark} = J_0 \exp(qV/(kT))\)  \cite{hirst2011fundamental, de1981thermodynamic}:

where \(J_0\) is given by, 
\begin{equation}
    J_0 = q  \int_{E_g}^{\infty} \frac{4\pi}{{c^2h^3}} E^2 \exp\left(\frac{-E}{kT}\right) dE
    \label{eq:J0}
\end{equation}
 \(V\) is the voltage, $T$ is the temperature, $k$ is the Boltzmann constant, $q$ is the charge of an electron, $h$ is Planck's constant, and $c$ is the speed of light in vacuum. We have used the AM1.5G spectrum as the incident solar spectrum. The photon flux above a certain bandgap is calculated by integrating the photon flux over the energy spectrum, starting from the energy corresponding to the bandgap to the maximum available energy. This determines the number of photogenerated carriers that contribute to the photonic current density, \(J_{ph}\). Thus, the effective current for a SJ solar cell is given by the difference in photonic and dark current densities, that is ($J_{ph} - J_{dark}$) \cite{strandberg2020analytic}: 
\begin{equation}
    J = J_{ph} - J_0 \cdot \exp{\left(V/V_t\right)}
    \label{single_junction_JV}
\end{equation}
where, \(V_t\) is the thermal voltage given by \(kT/q\).

For the $N$-layer CM tandem device configuration, shown in Fig. \ref{fig:device_structure}, each subcell can be considered as a SJ device. The current density in layer $i$ is given by:  
\begin{equation}
    J_{i} = \frac{J_{ph,i}}{m_i} - \frac{J_{0,i}}{m_i}\exp\left(\frac{V_i}{m_iV_t}\right) + \frac{J_{0,i+1}}{2m_{i}}\exp\left(\frac{V_{i+1}}{m_{i+1}V_t}\right)
    \label{eq:Ji_individual}
\end{equation}
Each individual subcell in layer \(i \) contributes a voltage of \({V_i}/{m_i} \) which gives a total voltage \( V_i \) for that layer. As the device is in series configuration, these voltages add up to give \(V\). As the photonic flux gets scaled, the photonic current density gets scaled as well. Similarly, the radiative coupling and dark current densities are scaled by the number of subcells in the corresponding layer.

Optimization of the number of subcells/modules in each layer requires equalizing the current densities at their maximum power point (\(MPP\)), \(J_{mpp}\) across all the layers, as the device is to have the same current across all layers \cite{strandberg2020analytic}. Thus, the general expression for $m_i$ in terms of the number of bottom-most cells ($m_1$) is given by the following equation:

\begin{widetext}
\begin{equation}
m_j = m_1 \cdot 2^{j-1} \left(  \frac{\psi\left\{ \left(\sum\limits_{i=j}^{N} 2^{N-i}\cdot J_{ph,i}\right), \left(2^{N-j}\cdot J_{0,j}\right)\right\}-\psi\left\{\left(\sum\limits_{i=j+1}^{N} 2^{N-i}\cdot J_{ph,i}\right), \left(2^{N-(j+1)}\cdot J_{0,j+1}\right)\right\}}{\psi\left\{\left(\sum\limits_{i=1}^{N} 2^{N-i}\cdot J_{ph,i}\right), \left(2^{N-1}\cdot J_{0,1}\right)\right\}-\psi\left\{\left(\sum\limits_{i=2}^{N} 2^{N-i}\cdot J_{ph,i}\right), \left(2^{N-2}\cdot J_{0,2}\right)\right\}} \right)
\label{eq: mi}
\end{equation}
\end{widetext}

For brevity, a function \(\psi\{x,y\}\) has been defined as follows 

\begin{equation}
    \psi\{x,y\} = x\cdot\left(1-\frac{1}{{W}\left(\frac{x}{y}\cdot e\right)}\right)
\end{equation}

where, \(W\) is the Lambert $W$ function \cite{corless1996lambert}.

On obtaining the optimal number of subcells across all the layers, an expression for the current density - voltage \(J\)-\(V\) for the complete tandem device can now be constructed. As the device is to be current-matched, the current densities across all layers are equated to \({J}\). On solving equation (\ref{eq:Ji_individual}) \(\forall i \in [1,N]\), a general equation for the \(J\)-\(V\) characteristics for an $N$-layer CM tandem device is obtained: 
\begin{equation}
    {V} = \sum\limits_{i=1}^{N} m_i V_t \ln\left( \frac{\left(\sum\limits_{j=1}^{i+1} 2^{i-j} \cdot J_{ph,j}\right) - \left(  \sum\limits_{j=1}^{i+1} 2^{i-j}\cdot m_j\right) {J}}{2^{N-i} \cdot J_{0,i}} \right)
    \label{eq:JV_tandem}
\end{equation}

\begin{figure*}[t]
    \centering
    \includegraphics[scale=0.8]{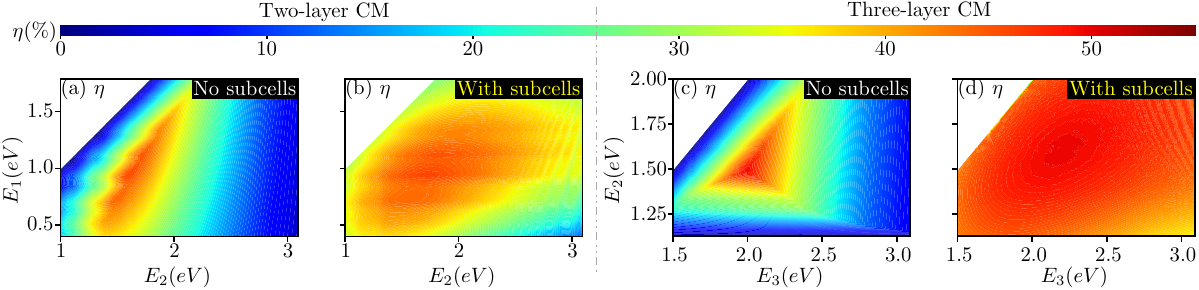}
        \caption{(a), (b): Efficiency of a two-layer CM tandem without and with subcells as a function of the top (\(E_2\)) and bottom (\(E_1\)) layer bandgaps. (c), (d): Efficiency of a three-layer CM tandem without and with subcells as a function of the top (\(E_3\)) and middle layer (\(E_2\)) bandgaps. The bottom-most layer in the three-layer configuration is \(Si\) with \(E_{1} = 1.12 eV\). The number of bottom-most subcells in both the tandems is taken to be $60$ ($m_1 = 60$). }
    \label{fig:results1}
\end{figure*}

\begin{figure*}[t]
    \centering
    \includegraphics[scale=0.8]{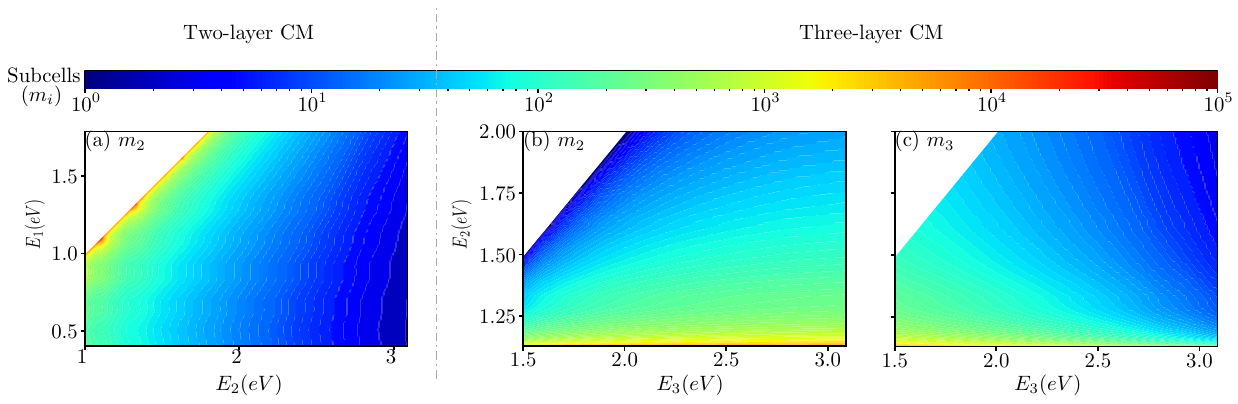}
    \caption{(a)-(c) Optimal number of subcells in CM tandem devices, with the color bar representing values on a logarithmic scale: (a) Optimal number of bottom subcells \(m_1\) in a two-layer CM tandem. (b) and (c) shows the optimal number of subcells for each layer in a three-layer CM tandem device: (b) shows the optimal number of subcells in layer \(1\) (\(m_1\)), and (c) in layer \(2\) (\(m_2\)). The bottom-most layer in the three-layer configuration is \(Si\) with \(E_{1} = 1.12 eV\). The number of bottom-most subcells in both the tandems is taken to be $60$ ($m_1 = 60$). }
    \label{fig:results2}
\end{figure*}

The voltage and current density at the MPP (\(V_{mpp}\) and \(J_{mpp}\)) can now be calculated from the \(J\)-\(V\) characteristics. Our framework is summarized in the flowchart shown in Fig. \ref{fig: flowchart}.

\section{Results and Discussion}
Our framework's capability is demonstrated by applying it to two-layer and three-layer CM tandem solar cells. Fig. \ref{fig:results1} summarises the results from our calculations for a two-layer [(a)-(b)] and three-layer CM tandem cell [(c)-(d)] considering configurations without and with subcells. For configurations that incorporate subcells, we assume the lowermost layer to contain \(60\) subcells (that is, \(m_1 = 60\)) Fig. \ref{fig:results2} shows the subcell counts for two-layer [(a)] and three-layer [(b)] CM tandems.

\begin{table*}[t]
\centering
\caption{Summary of maximum-efficiency CM tandem configurations (with and without subcells).}
\label{tab:cm-tandem-max-eff}
\begin{ruledtabular}
\begin{tabular}{l c l c}
\textbf{Configuration} & \textbf{Bandgaps (eV)} & \textbf{Subcell counts} & \textbf{Efficiency (\%)} \\
\hline
2-layer, no subcells    
  & $E_{1} = 0.94,\; E_{2} = 1.59$            
  & $m_{1} = 1,\; m_{2} = 1$  
  & 45.55 \\
2-layer, with subcells  
  & $E_{1} = 0.94,\; E_{2} = 1.73$            
  & $m_{1} = 60,\; m_{2} = 45$ 
  & 45.90 \\
\hline
3-layer, no subcells    
  & $E_{1} = 1.12,\; E_{2} = 1.49,\; E_{3} = 2.00$      
  & $m_{1} = 1,\; m_{2} = 1,\; m_{3} = 1$ 
  & 49.20 \\
3-layer, with subcells  
  & $E_{1} = 1.12,\; E_{2} = 1.64,\; E_{3} = 2.24$      
  & $m_{1} = 60,\; m_{2} = 45,\; m_{3} = 31$ 
  & 49.63 \\
\end{tabular}
\end{ruledtabular}
\end{table*}

\subsection{Two-layer Tandem}
Fig. \ref{fig:results1}(a) illustrates the efficiency (\(\eta\)) of a two-layer CM tandem device without subcells, that is, when \(m_2 = m_1 = 1\). It is observed that a small range of bandgap combinations of top and bottom layers yield high efficiencies, owing to the current-matching constraint imposed by the series connection of layers \cite{strandberg2015detailed}. 
%Additionally, it is discernible from Fig. \ref{fig:results1}(b) that the bandgap combinations that yield maximum \(\eta\) do not correspond to the ones that give maximum \(FF\); this is due to the latter's dependency on \(V_{mpp}, J_{mpp}, V_{oc}\), and \(J_{sc}\) \cite{boccard2020}. 

Fig. \ref{fig:results1}(b) shows $\eta$ plot for devices with subcells, where the optimal number of subcells is determined using equation (\ref{eq: mi}). The subcell counts are rounded off to the nearest natural number. Fig. \ref{fig:results2}(a) indicates the optimal number of subcells for the first layer. We can observe that as the bandgaps of both layers approach the same value, the top layer necessitates a significantly large number of subcells (See regions in Fig. \ref{fig:results2}(a) where \(E_2 = E_1\)). This is due to very little photon flux penetrating the bottom layer, leading to the generation of only a nominal amount of current density by this layer. Consequently, to match this low current density, the top layer requires a significantly large number of subcells in order to reduce its current output and match it to the bottom layer. From Fig. \ref{fig:results1} (b), we can observe that the spread of bandgap combinations that yield higher efficiencies has increased. This can be attributed to the alleviation of the current-matching constraint by incorporating the appropriate number of subcells in the layer with the lowest current density. The sharp edges in the contour plots are a consequence of the integer restriction imposed on the number of subcells.
%This is to be expected as we are, in essence, alleviating the constraint by adding the appropriate number of subcells to the layer with the lower current density. The jagged/sharp edges in the contour plot are due to the integer restriction imposed on the number of subcells.

It is well-known that unconstrained tandems have higher efficiency limits for a wider range of bandgaps \cite{strandberg2015detailed, hu2019}, compared to their constrained counterparts without subcells. As using subcells removed the current-matching constraint, the efficiency closely mirrors that of unconstrained tandems.
%The inclusion of subcells at each layer of a CM tandem device alleviates the limitations associated with current matching requirements, thereby enhancing performance akin to that of unconstrained counterparts.

While the maximum efficiency in a CM tandem without subcells is the same as that with subcells, there is a significant difference in the spread of bandgaps that yield high efficiencies. The range of combinations of bandgaps that yield high efficiencies is much larger in CM tandems with subcells [Fig. \ref{fig:results1}(b)], as compared to the one without [Fig. \ref{fig:results1}(a)]. This enhancement in design flexibility allows for the exploration of a wider array of bandgap combinations, thereby reducing the stringent constraints previously imposed on material choices.

A summary of the bandgaps and subcell counts that yield maximum efficiency for two- and three-layer CM tandems can be seen in Table \ref{tab:cm-tandem-max-eff}.

\begin{figure*}[t]
    \centering
    \includegraphics[scale=1]{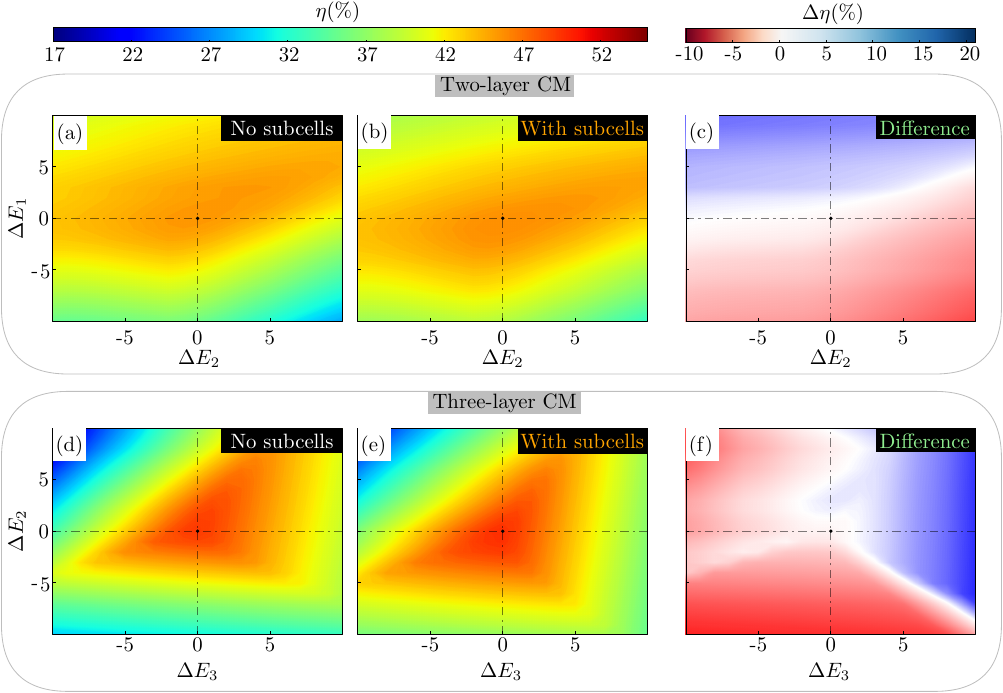}
\caption{(a),(b): Efficiencies of two-layer tandems without and with subcells, respectively, as a function of bandgap mismatch among layers. (d),(e): Efficiencies of three-layer tandems without and with subcells, respectively, as a function of bandgap mismatch among layers. (c),(f): Relative efficiency difference, defined as \(\Delta \eta = (1- \eta_\text{no subcells}/\eta_\text{subcells})\times 100\), for two-layer and three-layer CM tandems, respectively.}
\label{fig: mismatch}
\end{figure*}

% While the variation in maximum efficiency limits across different configurations is negligible, there is a significant difference in the selection of bandgaps that yield higher efficiencies. The range of combinations of bandgaps that yield high efficiencies is much larger in CM tandems with subcells [Fig. \ref{fig:results1}(c)], as compared to the one without [Fig. \ref{fig:results1}(a)]. This enhancement in design flexibility allows exploring a wider array of bandgap combinations, thereby reducing the stringent constraints previously imposed on material choices.

\subsection{Three-layer tandems}

Applying the framework to a three-layer CM tandem device reveals a similar trend. The bottom-most layer (Layer \(1\)) is assumed to be silicon, having a bandgap of $1.12eV$. The bandgaps of the top and middle layers are varied (Layer \(3\) and Layer \(2\)), and the performance limits are calculated accordingly. Fig. \ref{fig:results1}(c) shows the efficiency for the three-layer tandem. We can clearly see a smaller combination of bandgaps yielding maximum efficiencies without adding subcells, similar to the two-layer CM tandem.   
%As Si itself has a base performance limit, the overall efficiency starts at a relatively higher point in contrast to the two-layer CM tandem where all the bandgaps were varied [see Fig. \ref{fig:results1}(c) and (g)].  

Upon including subcells into that tandem [Fig. \ref{fig:results2} (d)], we see that the efficiency spread is uniform. As Si was chosen as the lowermost layer, due to Silicon's inherent baseline performance range, the overall efficiency initiates from a comparably elevated value, unlike in the two-layer CM tandem where all the bandgaps were varied [see Fig. \ref{fig:results1}(b) and (d)]. Analogous to the two-layer tandem, the inclusion of subcells enables broadening the range of bandgaps that yield high \(\eta\). It is seen that the region of maximum efficiency does not require as many subcells as regions with higher bandgaps [as seen in Fig. \ref{fig:results2}(b) and (c)]. This is due to the dependence of photonic current density on the bandgap; a subtle change in the bandgap is enough to maintain the current-matching constraint.  
Similar to the two-layer CM tandems, the maximum efficiency remains almost the same whether including subcells or otherwise. However, the flexibility in choosing bandgaps that yield high efficiencies has broadened. 

Though adding subcells does not increase the maximum possible efficiency, a wider range of bandgaps can be utilized, giving rise to high efficiencies. This can potentially enable the use of materials that typically have low efficiencies, but when integrated as area de-coupled subcells, can yield significantly higher efficiencies.

\subsection{Effects of Mismatch}\label{mismatch_shading}

Fig. \ref{fig: mismatch} shows the effects of bandgap mismatch on the efficiency of CM tandems with and without subcells. The bandgaps and subcells counts giving maximum efficiency (in accordance with Table. \ref{tab:cm-tandem-max-eff}) have been chosen. The center of each plot represents the ideal bandgap configuration, where there is no bandgap mismatch. A shift to the left or right along the \( X\)-axis corresponds to a decrease or increase in the bandgap value, respectively. Similarly, movement along the \( Y \)-axis follows the same trend, with downward shifts indicating a reduction in bandgap and upward shifts signifying an increase.

Fig. \ref{fig: mismatch} (a),(b) shows the efficiency of a two-layer CM tandem as a function of top-layer (\(\Delta E_2\)) and bottom-layer (\(\Delta E_1\)) bandgap mismatch for configurations without and with subcells, respectively. Fig. \ref{fig: mismatch} (c) shows the relative efficiency of tandems with subcells with respect to no subcells, \(\Delta \eta = (1- \eta_\text{no subcells}/\eta_\text{subcells})\times 100\). Similarly, Fig. \ref{fig: mismatch} (d),(e) shows the efficiency of a three-layer CM tandem as a function of the top layer (\(\Delta E_3\)) and middle layer (\(\Delta E_2\)) bandgap mismatch for configurations without and with subcells, respectively. Fig. \ref{fig: mismatch} (f) shows the relative efficiency of tandems with subcells with respect to no subcells.

We can see that for both two-layer and three-layer tandems, both configurations (without and with subcells) achieve high efficiencies near zero bandgap mismatch, located around the center of the plots. However, we see that moving away from \(0\) mismatch, CM tandems with subcells maintain higher efficiencies compared to their no-subcell counterparts, as can be seen from Fig. \ref{fig: mismatch} (c) and (f). In the case of two-layer tandems, for high mismatch in bottom layer bandgaps, we see that subcell configuration yields lesser efficiency compared to no subcell configurations. This is due to the fixed subcell counts. If \(m_2\) were adjusted to the new bandgaps caused due to the mismatch (as per Fig. \ref{fig:results2}(a)), the subcell configuration would match or exceed the no-subcell efficiency. A similar phenomenon can be observed in the case of a three-layer CM tandem (see Fig. \ref{fig: mismatch} (f)).

\section{Scope and Limitations}\label{Scope and Limitations}
The purpose of this paper was to ascertain the fundamental thermodynamic limits of efficiency for current-matched (CM) tandem solar cells that incorporate subcells. Our framework is based upon the principle of detailed balance and part of the assumptions associated with it was that radiative recombination was the only loss mechanism considered. Series and shunt resistances can affect device performance, with series resistance leading to ohmic losses in carrier transport layers and shunt resistances responsible for capturing leakage paths which lowers open-circuit voltage \cite{VanDyk_Meyer_2004,Barbato_Meneghini_Cester_Mura_Zanoni_Meneghesso_2014,Baba_Makita_Mizuno_Takato_Sugaya_Yamada_2018}. These resistive losses impact both configurations with and without subcells. However, the ability to tune the number of subcells offers the advantage of using a wider range of bandgaps that yield high efficiencies, as seen in Fig. \ref{fig:results2}. This tunability ensures that subcell configurations maintain superior performance, even when accounting for the effects of series and shunt resistances. We have demonstrated the framework's capability by simulating two-layer and three-layer tandems but it can be easily extended to \(N\)-layer tandems. The model serves as a general framework for the analysis of efficiency limits and provides design insights into the number of subcells of any number of layers and can be extended to voltage-matched (VM) devices and bifacial devices as well.

\section{Conclusion}\label{Conclusion}
In this work, we have presented an analytical framework that calculates the performance limits of an \(N\)-layer current-matched (CM) tandem solar cell with and without subcells at each layer, accounting for radiative coupling between the layers. The inclusion of subcells across layers in the CM tandem configuration, while not enhancing overall efficiency, significantly broadened the range of bandgaps that achieve high efficiencies, with performance limits similar to their unconstrained counterparts. An expression for the optimal number of subcells for each layer has been derived, and the performance limits for two-layer and three-layer CM tandem devices have been presented as an example. We have also shown that in the presence of mismatched bandgaps, CM tandems with subcells tend to maintain higher efficiencies compared to configurations without subcells. Our work provides useful design insights into the selection of subcells in each layer for any CM tandem configuration; this approach can be extended to include voltage-matched and bifacial tandems as well.

% Use the style of numbering in square brackets.
% If nothing is used, default style will be taken.
%\begin{enumerate}[a)]
%\item 
%\item 
%\item 
%\end{enumerate}  

%\begin{itemize}
%\item 
%\item 
%\item 
%\end{itemize}  

%\begin{description}
%\item[]
%\item[] 
%\item[] 
%\end{description}  

% Figure
% \begin{figure}[<options>]
% 	\centering
% 		\includegraphics[<options>]{}
% 	  \caption{}\label{fig1}
% \end{figure}

% \begin{table}[<options>]
% \begin{tabular*}{\tblwidth}{@{}LL@{}}
% \toprule
% \midrule
%  & \\
%  & \\
%  & \\
%  & \\
% \bottomrule
% \end{tabular*}
% \end{table}

% Uncomment and use as the case may be
%\begin{theorem} 
%\end{theorem}

% Uncomment and use as the case may be
%\begin{lemma} 
%\end{lemma}


%

% \nocite{*}
\bibliography{aipsamp}% Produces the bibliography via BibTeX.

\end{document}